\documentstyle[graphics,aps,amssymb]{revtex}

\newcommand{\bfg}[1]{\mbox{\boldmath $#1       $}}

\begin{document}

\title{Three-dimensional numerical simulations of 1 GeV/Nucleon U$^{92+}$
impact against atomic hydrogen}

\author{J. R. V\'azquez de Aldana and Luis Roso}

\address{ Departamento de F\'\i sica Aplicada, Universidad de Salamanca,
E-37008 Salamanca, Spain}

\date{\today}
\maketitle

\begin{abstract}

The impact of 1 GeV/Nucleon U$^{92+}$ projectiles against atomic hydrogen is
studied by direct numerical resolution of the time-dependent non-relativistic
wave-equation for the atomic electron on a three-dimensional Cartesian lattice.
We employ the fully relativistic expressions to describe the electromagnetic fields created by
the incident ion.
The wave-equation for the atom interacting with the projectile is carefully derived from the
time-dependent Dirac equation in order to retain all the relevant terms.

\end{abstract}

\pacs{34.10.+x, 34.50.Fa}

\section{Introduction}

Atomic ionization by an electromagnetic wave, either incoherent or
coherent, has been widely studied since the old times of
Quantum Mechanics in many different situations.
It is also obvious, but much less common in this context, to study ionization by the
electromagnetic fields generated by a fast charged projectile passing
nearby.
In the case of a laser field, or any other electromagnetic wave, the field is a radiation
field far from the sources, in other words, it is made of real photons \cite{Protopapas97}.
However in case of a rapidly passing projectile, the field has different properties and it is just made of
virtual photons.

That those virtual photons can ionize the atom is completely clear, but the
ionization dynamics should be quite different from the laser field case.
Recently, considerable interest has been devoted to this subject \cite{ion-foton},
both for single and multiple ionization of atoms with fast ions.
However, very few experiments have been reported with relativistic incident ions.
At the GSI in Darmstadt \cite{Ullrich97}, experiments with relativistic incident ions
have been performed. The GSI experiment studied the single and double
ionization of helium by 1 GeV/Nucleon U$^{92+}$ impact. As these authors state, the
relativistic ion generates a sub-attosecond super-intense electromagnetic pulse.

Due to the geometry of the system (incident projectile and nucleus), there are no
symmetries present except for a mirror symmetry along the plane defined by the
projectile trajectory and the initial position of the nucleus. Therefore three-dimensional
studies are needed and this is extremely difficult to be done {\it ab initio} for a
two electron system.
In the present paper we present a very realistic description of
the (relativistic) ion-atom interaction but just for one-electron atoms.
We compute the electron wave-function in a Cartesian three-dimensional
lattice, taking into account the relativistic dynamics of the projectile.

Calculations in three-dimensional Cartesian grids have been shown to
be a very good description for the $p+H$ collisions \cite{proton-hyd},
with non-relativistic incident protons. The agreement with experimental results
is remarkably good for the range of parameters considered.
Collisions with antiprotons have been also investigated with these {\it ab initio}
simulations \cite{antiproton-hyd,antiproton-he}.
However these calculations are done in a context that is not able to describe
the peculiar features of the electromagnetic fields (electric and magnetic fields)
generated by the relativistically moving projectile.
To our knowledge no previous studies have been done based on
the numerical direct resolution of the wave-equation, in cases where
the incident ion is relativistic.

A remarkable side consequence has been found. We need to describe a Lorentz
transformed Coulomb field instead of the typical radiation fields studied
in photo-ionization.
For the correct quantum description of such interaction, it is necessary to
include one extra term in the time-dependent Schr\"odinger equation.
This term is missing in most of the studies on laser photo-ionization, that
start from a time-dependent Schr\"odinger equation conceptually wrong.

The paper is organized as follows. In Sec. II we describe the electromagnetic fields
created by a relativistically moving ion. In our calculations, we neglect both the
recoil of the hydrogen nucleus so as the change in the trajectory of the ion:
these approximations are justified in terms of classical simulations (Monte Carlo)
in Sec. III. The derivation and discussion of the time-dependent wave-equation
for the atomic system is described in Sec. IV. A detailed description of the
integration algorithms and numerical techniques is also given.
In Sec. V we establish the parameters of the calculations and we present the
obtained numerical results.
The final Section is devoted to conclusions.

\section{Electromagnetic fields}

As our particular choice of the coordinates, to simplify the notation without loss of
generality, we consider that the heavy ion projectile is moving in the plane $z=0$
in parallel direction to the x-axis. The hydrogen nucleus is initially at the origin and the
projectile trajectory is $x_{ion}(t)=x_0+v t = x_0+\beta c t $, $y_{ion}(t)=b$
and $z_{ion}(t)=0$.
$b$ is the impact parameter (minimum distance to the nucleus in the trajectory of the ion).
The notation $\beta =v/c$ is standard in relativity.
The projectile is so heavy ($M_{ion}=238m_N$, being $m_N$ the nucleon mass) and
so energetic that we can perfectly assume that its trajectory is not affected by the atom.
This point is discussed in Sec. III in terms of classical simulations.

It is not difficult to show \cite{Jackson} that the electric and magnetic
fields created by such a relativistic projectile at an arbitrary point $(x,y,z)$ of space are,
\begin{eqnarray}
{\bf E}({\bf r},t)={{\,Z\, e \gamma} \over {\left[ {\gamma ^2\left(
{x-x_0-vt}
\right)^2+(y-b)^2+z^2}
\right]^{3/2}}}\;\left[ {\left( {x-x_0-vt} \right),y,z} \right]
\end{eqnarray}
and
\begin{eqnarray}
{\bf B}({\bf r},t)={{-\,Z\,e\gamma ^3\quad v/c} \over {\left[ {\gamma
^2\left({x-x_0-vt} \right)^2+(y-b)^2+z^2} \right]^{3/2}}}\;\;
\left( {0,z,-y} \right) \, ,
\end{eqnarray}
with $\gamma=1/\sqrt{1-\beta^2}$.
$Z$ indicates the projectile ion charge, that for the case studied here is $Z$=92.

These fields are generated by a scalar potential
\begin{eqnarray}
\eta \left( {\bf r},t \right)={{\,Z\,e\gamma } \over {\sqrt {\gamma
^2\left( {x-x_0-vt}
\right)^2+(y-b)^2+z^2}}}
\label{eq:escalar}
\end{eqnarray}
and a vector potential
\begin{eqnarray}
{ \bf A}\left( {\bf r},t \right)={{\,Z\,e\gamma \,{ \bf
v}/c} \over {\sqrt
{\gamma ^2\left( {x-x_0-vt} \right)^2+(y-b)^2+z^2}}} \, .
\label{eq:vector}
\end{eqnarray}
The velocity vector has only one component, ${ \bf v}=(v,0,0)$. Therefore,
${\bf A}({\bf r},t)=A_x\left( {\bf r},t \right) {\bf e}_x$ and $A_y=A_z=0$.

\section{Classical simulation}

Although we give, in this paper, a quantum description of the dynamics of the atomic electron interacting
with the ion, it is worth to start with a classical simulation of the three-body problem studied.
For the considered parameters of the incident ion, it is almost
trivial to perform classical Monte Carlo simulations for the motion of the
projectile ion, the proton and the electron. The goal of such simulations is just to
introduce the right approximations for the subsequent quantum description.

A few conclusions appear from the classical simulations for
heavy projectiles (U$^{92+}$) that move relativistically (1 GeV/Nucleon, or more).
First, the projectile trajectory is not changed at the space scale and with the precision
that we are interested in.
Second, the hydrogen nucleus is accelerated after the collision with the
projectile but the momentum transfer is negligible unless we consider impact parameters
smaller than 0.1 a.u.
We are not interested in head-on collisions, that could be relevant
for nuclear physics.

\begin{figure}
\begin{center}
\scalebox{0.65}{\rotatebox{180}{\includegraphics{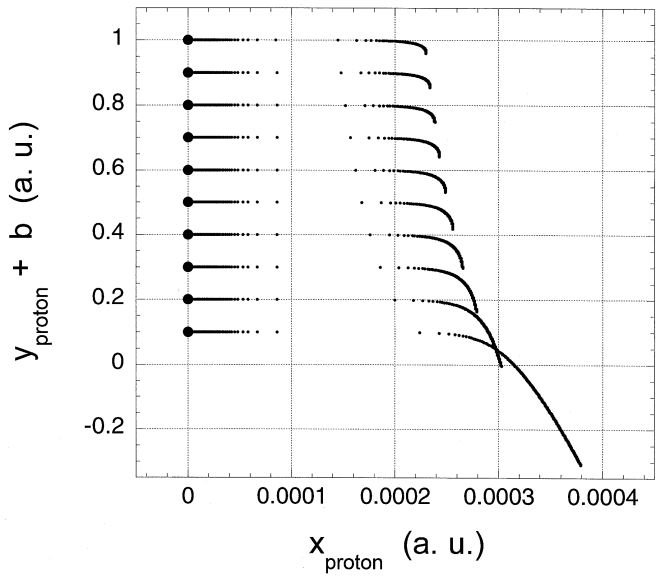}}}
\end{center}
\caption{Classical trajectories of the atomic nucleus for different values of the impact parameter $b$.
The horizontal axis represents the motion in the $x$ direction and the vertical axis represents
the motion in the $y$ direction, where the impact parameter has been added for better observing
the different trajectories. Distances are in atomic units.}
\label{fig:recoil}
\end{figure}

In Fig. \ref{fig:recoil} we represent the motion of the hydrogen nucleus
for different values of the impact parameter ($b$). The horizontal axis shows
the motion in the longitudinal direction ($x$ coordinate), and the vertical axis
shows the motion in the transversal direction ($y$ coordinate), where the impact parameter
has been added for better observing all the curves. Both scales are in atomic units.
The trajectories has been plotted since the incident projectile is placed at
$x(t)=-10$ a.u. (marked with big circles) until it reaches $x(t)=10$ a.u.

A result is clear: for the study of the ionization up to the level of
accuracy we are
interested in, it is fairly exact to consider that the projectile ion
follows its trajectory
unaltered. It is also reasonable, due to the short interaction time, to
consider that the
hydrogen nucleus remains unaltered at its initial position. Therefore only
the electron motion needs to be described in detail.
On the other side, the classical simulations indicate
that the electron's motion is essentially non-relativistic, except for
extremely small impact parameters.

\section{Time-dependent wave-equation}

Because the electron is interacting with such time and space dependent scalar and
vector potentials, it is worth to carefully describe the structure of the wave-equation.
It is well known that the Dirac equation for an electron interacting
with an arbitrary electromagnetic field ${\bf E}({\bf r},t)$ and ${\bf B}({\bf r},t)$ described with
the scalar potential, $\eta \left( {{\bf r},t} \right)$, and the vector potential,
${\bf A}\left( {{\bf r},t}\right)$,
can be written as a second-oder differential equation \cite{Schiff}
\begin{eqnarray}
\left( {i\hbar {\partial  \over {\partial t}}+e\eta +e V}
\right)^2 \Phi_D =\left( {c{\bf p}+e{\bf A}} \right)^2\;\Phi_D +m^2c^4\;\Phi_D
+ e\hbar c\;\bfg{\sigma}' \cdot {\bf B}\;\Phi_D- i e\hbar c\; \bfg{\alpha} \cdot
{\bf E}\;\Phi_D \, ,
\label{dirac}
\end{eqnarray} where $\Phi_D$ is the standard Dirac four-spinor, and
$\bfg{\sigma}'$ and $\bfg{\alpha}$ are $4\times 4$ matrices.
The space and time dependence on the fields, potentials and wave-function has been removed for
simplicity in the expressions. To describe the Coulomb potential of the hydrogen atom we
have also included the $eV$ term.

To move to the non-relativistic domain we start by neglecting the
$-ie\hbar c\; \bfg{\alpha} \cdot {\bf E} \; \Phi_D $ term, as it
mixes particle and antiparticle states.
We also neglect the coupling between spin and magnetic field,
$\bfg{\sigma}'\cdot {\bf B}$, because the degree of freedom of
spin is not expected to play an important role in the dynamics of the electron,
in the cases that we are considering.
Now we can just consider a one-component wave-function $\Phi$
\begin{eqnarray}
\left( {i\hbar \frac{\partial}{\partial t}+e\eta+eV }
\right)^2 \Phi =\left( {c{\bf p}+e{\bf A}} \right)^2\;\Phi +m^2c^4\;\Phi \, .
\end{eqnarray}
Expanding the square that includes the time derivative, and taking into
account that {\it the scalar potential can be time-dependent}:
\begin{eqnarray}
-\hbar ^2\; \frac{\partial ^2 \Phi}{\partial^2 t}
+2ie\hbar (\eta+V) \frac{\partial \Phi}{\partial t} +
ie\hbar \left( \frac{\partial \eta }{\partial t} \right) \Phi
+e^2(\eta+V) ^2\;\Phi   =\left( { c{\bf p}+e{\bf A}} \right)^2\;\Phi  +
m^2c^4\;\Phi \, .
\end{eqnarray}
We now introduce explicitely the fast time oscillation,
\begin{eqnarray}
\Phi  \left( {\bf r},t \right)= \Psi \left( {\bf r},t \right)\;\exp
\left( -i\;\frac{mc^2}{\hbar }t \right)
\end{eqnarray}
 and the wave-equation becomes,
\begin{eqnarray}
&&2i\hbar mc^2\frac{\partial  \Psi }{\partial t}
-\hbar ^2\frac{\partial ^2 \Psi }{\partial ^2 t}
+2 e m c^2 (\eta+V) \; \Psi+2i\;\hbar \;e(\eta+V) \frac{\partial \Psi }{\partial t}
\nonumber\\
&&+i\;e\hbar \;\frac{\partial \eta}{\partial t} \; \Psi
+e^2(\eta+V) ^2\; \Psi
=\left( { c{\bf p}+e{\bf A}}
\right)^2\; \Psi
\end{eqnarray}
so that we have eliminated the fast oscillation due to the mass term.
The next step towards the non-relativistic wave-equation consists of neglecting the
$\partial ^2 \Psi  / \partial ^2t$ term, because the fast oscillations due
to the mass term have been explicitly accounted for.
Moreover it is consistent with the non-relativistic limit we are looking for to assume,
$m\,c^2 >> e (\eta+V) $.  Thus the wave-equation is simplified to,
\begin{eqnarray}
2i\hbar mc^2 \; \frac{\partial \Psi}{\partial t}
 +\left[ {2emc^2(\eta+V)+ie\hbar  \frac{\partial \eta }{\partial t}} \right] \Psi
 =\left( { c{\bf p}+e{\bf A}}
\right)^2\; \Psi \, .
\end{eqnarray}
The $i e\hbar  \partial \eta / \partial t$ term {\it should not} be eliminated
because even when the modulus is very small compared with the electron's rest
mass, it is dephased due to the $i$ factor.
Therefore it is necessary to keep this term,
\begin{eqnarray}
i \hbar \frac{\partial  \Psi}{\partial t}=
\frac{1}{2m}\left( {\bf p}+\frac{e}{c} {\bf A} \right)^2 \Psi
- e \eta \; \Psi - e V \; \Psi - i \frac{e\hbar}{2mc^2} \frac{\partial \eta}{\partial t}\; \Psi \, .
\end{eqnarray}
It is worth to write this wave-equation in the form of a Hamiltonian system.
This Hamiltonian has an extra term $i \partial\eta / \partial t $,
that needs to be included in the time-dependent Schr\"odinger equation.
This term is, however,
fundamental for the validity of this non-relativistic equation for any
scalar and vector
potentials with arbitrary space-time dependences.

Expanding the $\left( {\bf p}+e/c{\bf A}\right)^2$ term, and taking care
of the non-commutability of ${\bf p}$ and ${\bf A}$, one gets,
\begin{eqnarray}
i \hbar \frac{\partial \Psi }{\partial t}=\frac{{\bf p}^2}{2m}\; \Psi
+\frac{e}{mc} {\bf A}\cdot {\bf p} \; \Psi
+\frac{e^2}{2mc^2} {\bf A}^2  \; \Psi -
e\eta \; \Psi -eV \; \Psi -i\frac{e\hbar}{2mc}\left( {\bfg{\nabla} \cdot {\bf A}+\frac{1}{c} \frac{\partial \eta}{\partial t}}
\right)\Psi \, .
\end{eqnarray}
If we introduce the Lorentz condition for the scalar and vector potentials,
\begin{eqnarray}
\bfg{\nabla} \cdot {\bf A}+\frac{1}{c} \frac{\partial \eta}{\partial t}=0
\end{eqnarray}
the wave-equation is heavily simplified,
\begin{eqnarray}
i\hbar \frac{\partial  \Psi }{\partial t}
=\frac{{\bf p}^2}{2m} \; \Psi +\frac{e}{mc} {\bf A}\cdot {\bf p} \; \Psi
+\frac{e^2}{2mc^2} {\bf A}^2  \; \Psi -e\eta\; \Psi -e\eta\; V \, .
\label{eq:TDE}
\end{eqnarray}

Observe that this equation is different from the standard time-dependent Schr\"odinger equation,
that considers the term $\left( { c{\bf p}+e{\bf A}} \right)^2$: it includes the divergence of the
vector potential but it does not include the very important $i \partial \eta  / \partial t$ term.
Such equation is not general and can not be safely employed to describe the interaction with
arbitrary electromagnetic fields. Only transversal electromagnetic fields can be accurately described.
We have introduced this derivation prior to start computing the dynamics of the electron
in order to clarify this point, that could become a source of error in the description of the system.

One final remark on the above equation. The fields that we describe with ${\bf A}$ and $\eta$,
Eq. (\ref{eq:escalar}) and Eq. (\ref{eq:vector}), are invariant under Lorentz transformations.
The time-dependent Schr\"odinger equation is invariant under Galilean transformations
due to its non-relativistic nature.
However, Eq. (\ref{eq:TDE}) shows neither one nor the other kind of invariance.
This formal inconsistency in our model is not important if the dynamics of the
electron is non-relativistic (like in the cases that we are considering), and thus
Lorentz transformations reduces to Galilean transformations (in the low velocity regime).

The final equation that we consider is:
\begin{eqnarray}
i\hbar\frac{\partial \Psi({\bf r},t)}{\partial t}&=&-\frac{\hbar^2}{2m}
\bfg{\nabla}^2 \Psi({\bf r},t)-
i\frac{e\hbar}{mc}{\bf A}({\bf r},t)\cdot\bfg{\nabla} \; \Psi({\bf r},t)
\nonumber\\
&+&\frac{e^2\hbar^2}{2mc^2}{\bf A}({\bf r},t)^2 \; \Psi({\bf r},t)
-e\eta({\bf r},t)\Psi({\bf r},t) -eV({\bf r}) \; \Psi({\bf r},t)  \, .
\label{eq:tdse}
\end{eqnarray}
$V({\bf r})=e/\sqrt{x^2+y^2+z^2}$ is the Coulomb potential.
The electromagnetic potentials are given by Eqs. (\ref{eq:escalar}) and (\ref{eq:vector}),
and they already verify the Lorentz condition because they come from the
Lorentz transform of a Coulomb potential.

We numerically solve Eq. (\ref{eq:tdse}) on a three-dimensional Cartesian
lattice with a uniform grid spacing $\Delta x=\Delta y=\Delta z=0.3$ a.u.
The lattice is taken in such a way that avoids the singularity
in the origin by considering grid points at half integer times the grid spacing.

Our numerical technique is based on a symmetric splitting of the
time-evolution operator
so that the wave-function at a given time $t+\Delta t$ is calculated from
the wave-function
in the previous time step $t$:
\begin{eqnarray}
\Psi({\bf r},t+\Delta t) &\simeq&
\exp \left[ -i\frac{\Delta t}{\hbar} {\hat H}(t+\Delta t/2)\right]
\Psi({\bf r},t)
\nonumber\\ & \simeq & \exp \left[ -i\frac{\Delta t}{2\hbar} {\hat
H}_z(t+\Delta t/2)
\right]
\exp \left[ -i\frac{\Delta t}{2\hbar} {\hat H}_y(t+\Delta t/2) \right]
\nonumber\\ &
\times & \exp \left[ -i \frac{\Delta t}{\hbar} {\hat H}_x(t+\Delta t/2) \right]
\exp \left[ -i\frac{\Delta t}{2\hbar} {\hat H}_y(t+\Delta t/2) \right]
\nonumber\\ &
\times & \exp \left[ -i\frac{\Delta t}{2\hbar} {\hat H}_z(t+\Delta t/2)
\right]  \Psi({\bf
r},t) + O(\Delta t^3)     \, .
\end{eqnarray} We have defined the hamiltonian operators as follows:
\begin{eqnarray} {\hat H}_x &=&-\frac{\hbar^2}{2m}
\frac{\partial^2}{\partial x^2}
+\frac{e^2\hbar^2}{2mc^2}{\bf A}({\bf r},t)^2 -
i\frac{e\hbar}{mc}A_x({\bf r},t)\frac{\partial}{\partial x}- e\eta({\bf
r},t)-eV({\bf r})
\nonumber\\ {\hat H}_y &=&-\frac{\hbar^2}{2m} \frac{\partial^2}{\partial
y^2} \, ,  \,
{\hat H}_z =-\frac{\hbar^2}{2m} \frac{\partial^2}{\partial z^2} \, .
\end{eqnarray}
The exponentials are thus expressed in the Cayley unitary
form \cite{Recipes}, that preserves the norm of the wave-function, and a Crank-Nicholson scheme is
employed to evaluate the space derivatives.
Similar methods have been successfully employed to integrate the non dipole
Schr\"odinger equation for the interaction of atoms with very intense laser fields \cite{JaviLuis01}.

The initial state for our calculations (the ground state $\phi_{1s}(r)$ of atomic
hydrogen) is computed in a $100\times 100 \times 100$ lattice with
the imaginary time propagation method. The computed energy for this state is $E_B=-0.490$ a.u.

Two different grid sizes have been employed to describe the ion-hydrogen collision,
depending on the value of the impact parameter ($b$):
\begin{eqnarray} N_x\times N_y \times N_z =
\cases{ 280 \times 380 \times 380 & for $b<10$ a.u. \cr 280 \times 280 \times
280 & for $b \geq 10$ a.u.}
\end{eqnarray}
Absorbers (a mask function) were employed at the integration
boundaries
in order to avoid reflections of the ionized population. The mask function
has the form $\sin^{1/8}$ and it is applied over $40$ points along the edge of the grid.
However, the interaction time is short enough to avoid a very large amount of population
reaching the boundaries (absorbed population is always smaller that $1$ percent).

\section{Numerical results}

In our calculations, the hydrogen nucleus is placed at the origin of the Cartesian coordinates.
For the selected energy of the incident ion (1 GeV/Nucleon) the relativistic parameters are
$\beta=0.3696$ and $\gamma=1.076$.
The U$^{92+}$ nucleus is initially at $x(0)=x_0=-200$ a.u., $y(0)=b$ and $z(0)=0$ a.u.
That choice of the initial value of $x(0)$ is a good compromise between accuracy -due to the long
range of the Coulomb potential- and a reasonable computer time.
The electronic wave-function is propagated in time until the ion reaches
$x(\tau)=600$ a.u. ($\tau=15.8$ a.u. is the final time). $2667$ time
iterations are employed, so that the ion covers a distance equal to the grid spacing
($0.3$ a.u.) each time step ($\Delta t = \Delta x /v = 0.00592$ a.u.).
In order to avoid problems with the singularity of the Coulomb potential of the ion,
the projectile moves equidistantly to the nearest points of the lattice.

We investigate the effect of the impact parameter in the dynamics of the atomic electron
by varying $b$ (non-uniformly) from $1$ a.u. to $150$ a.u.
We avoid smaller impact parameters because they
contribute negligibly to electron ionization cross sections and they turn
out to be relevant for ion-proton scattering (a physical situation out of the scope of the
present paper).

We evaluate, during the time propagation, the projection of the wave
function over the ground state $\phi_{1s}(r)$ to compute the total excitation probability
(probability of finding population both in free states as in bound excited states), for
each value of the impact parameter:
\begin{equation}
P_{exc}(b,t)=1- \left| \int_V{\Psi({\bf r},t)
\phi_{1s}(r)  d^3{\bf r}}
\right|^2 \, .
\label{eq:pexc}
\end{equation}
$V$ is the volume of the integration box.

\begin{figure}
\begin{center}
\rotatebox{270}{\scalebox{0.4}{\includegraphics{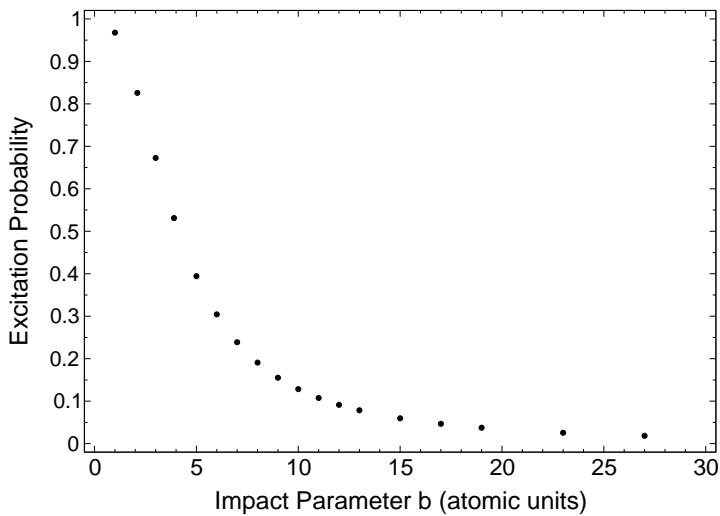}}}
\end{center}
\caption{Excitation probability $P_{exc}$ [see Eq. (\ref{eq:pexc})] at the final time $t=\tau$,
in terms of the impact parameter $b$.}
\label{fig:pexc}
\end{figure}

In Fig. \ref{fig:pexc} it is represented the computed values of $P_{exc}$
for different impact parameters $b$, at the final time $t=\tau$. In fact, due to the
fast motion of the ion, very few iterations after the ion has passed at the minimum
distance to the nucleus ($b$),
the excitation probability remains approximately unchanged. From
these values it is possible to estimate the total excitation cross section at the final
time, defined as
\begin{equation}
\sigma_{exc}(\tau)=2\pi \int_0^\infty {P_{exc}(b,\tau)b db} \, .
\end{equation}
We interpolate the excitation probability at the
intermediate points with
cubic-spline standard methods. The obtained value is $\sigma_{exc}(\tau)\simeq 8.4 \times
10^{-15} $ cm$^2$.

\begin{figure}
\begin{center}
\rotatebox{270}{\scalebox{0.4}{\includegraphics{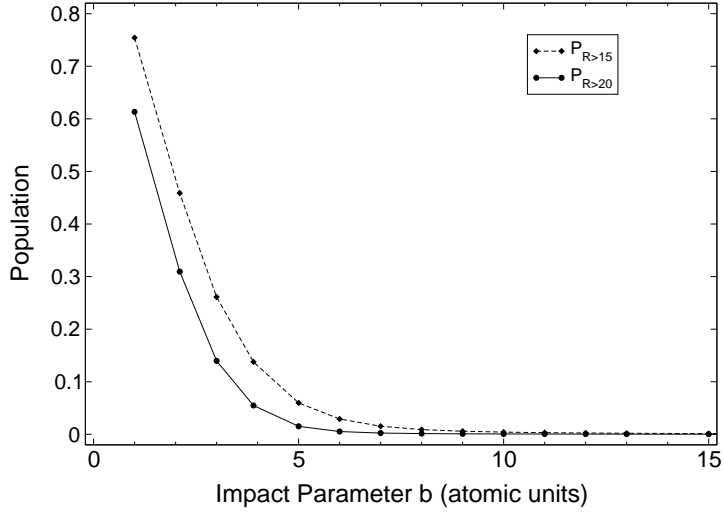}}}
\end{center}
\caption{Probability of finding population at distances greater than $R=15$
a.u. from the nucleus ($P_{R>15}$, in dashed lines and diamonds) and greater
that $R=20$ a.u. ($P_{R>20}$ in solid lines and circles), at the final time $t=\tau$.}
\label{fig:p1520}
\end{figure}

To gain insight into the dynamics of the electrons, we compute the
population that
remains, at any time of the interaction, at a distance $R$ to the nucleus
greater than
$15$ a.u. [$P_{R>15}(t)$] or greater than $20$ a.u. [$P_{R>20}(t)$], that
can be regarded as ionized electrons.
At the final time $\tau$, both values of the population have been plotted in Fig.
\ref{fig:p1520} in terms of
the impact parameter $b$. Dashed lines and diamonds corresponds to
$P_{R>15}(\tau)$
while solid lines and circles correspond to $P_{R>20}(\tau)$.
Both curves sharply decreases from $b=1$ to $b=5$ a.u., so that very few
electrons can be measured for larger impact parameters.

\begin{figure}
\begin{center}
\rotatebox{270}{\scalebox{0.43}{\includegraphics{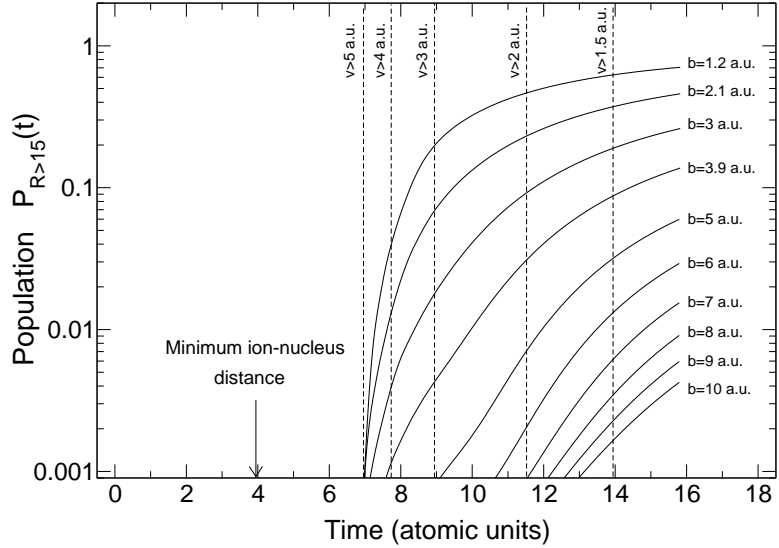}}}
\end{center}
\caption{Population at distances greater than $R=15$ a.u. during
the interaction time ($P_{R>15}(t)$), for impact parameters
from $1.2$ to $10$ a.u.
An approximated spectrum of the ejected electrons
can be estimated for different values of the impact parameter.
See text for explanation.}
\label{fig:spc}
\end{figure}

In Fig. \ref{fig:spc} we plot the time evolution of the population $P_{R>15}(t)$,
for different values of the impact parameter.
The horizontal axis contain the interaction time in atomic units
and the vertical axis the population $P_{R>15}(t)$.
We may approximately know the velocity distribution of the ejected
electrons, by simply
assuming that ionization process occurs at the time when the ion is placed
at the minimum distance to the nucleus [$x_{ion} (t_i)=0$, $y_{ion} (t_i)=b$ and
$z_{ion} (t_i)=0$].
In fact this is a good approximation due to the very high velocity of the projectile.
The time $t_i$ at which the ion is placed at this minimum distance
is marked with a vertical arrow.
Vertical dashed lines indicate the minimum necessary velocity of the
ejected electrons, $v_m$, to reach the distance $R=15$: $v_m(t) \simeq (t-t_i)/15$ a.u. (for $t>t_i$).
Observe that no electrons are ejected with velocities greater that $5$ a.u.
($v=0.037c$). Thus, to employ the non-relativistic wave-equation is justified
to describe the electron dynamics.

\begin{figure}
\begin{center}
\scalebox{0.65}{\includegraphics{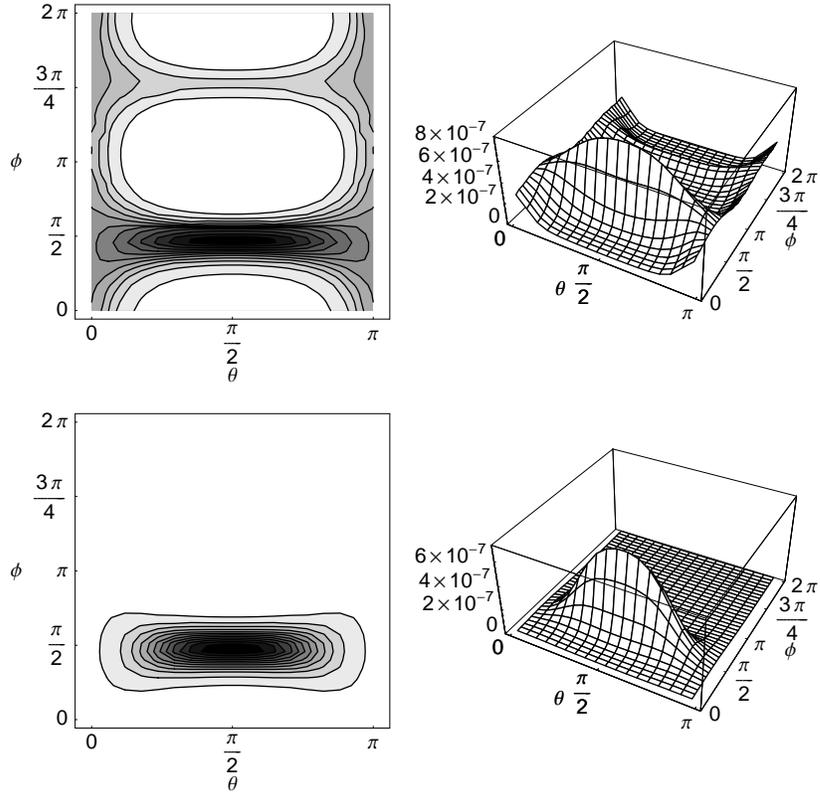}}
\end{center}
\caption{Angular distribution of the ejected electrons for the impact parameter
$b=1$ a.u. (plots at the top) and $b=3$ a.u. (plots at the bottom).
The plots show the population that at the final time is contained between
$R=15$ a.u. and $R=30$ a.u. Contour lines are in linear scale, with dark regions
indicating more populated regions.}
\label{fig:ang-dist}
\end{figure}

At the final time $\tau$, we have also computed the angular distribution of
the electrons.
To do so we consider electrons placed between two spheres, i.e. electrons
placed at a distance R from the nucleus such that
$R_1<R<R_2$ with $R_1=15$ a.u. and $R_2=30a.u.$ These values of the radii mean that we are
considering
electrons ejected with approximated velocities in the range $1.3<v_m<2.6$
a.u. In Fig.
\ref{fig:ang-dist} this angular distribution has been plotted for the
impact parameter
$b=1$ a.u. (plots at the top) and $b=3$ a.u. (plots at the bottom). The
angular
distribution per volume unit is defined by:
\begin{equation}
p(\theta,\phi)=\frac{P(R_1<R<R_2,\theta,\phi)}{\Omega(R_1<R<R_2,\theta,\phi)} \,
,
\end{equation} where we have defined:
\begin{eqnarray} P(R_1<R<R_2,\theta,\phi)&=&\int_{R_1}^{R_2}
\int_{\theta-\Delta
\theta/2}^{\theta+\Delta \theta/2}
\int_{\phi-\Delta \phi/2}^{\phi+\Delta \phi/2} |\Psi(r,\theta,\phi)|^2 r^2
\sin \theta dr
d\theta d\phi \nonumber\\
\Omega(R_1<R<R_2,\theta,\phi)&=&\int_{R_1}^{R_2} \int_{\theta-\Delta
\theta/2}^{\theta+\Delta \theta/2}
\int_{\phi-\Delta \phi/2}^{\phi+\Delta \phi/2} r^2 \sin \theta dr d\theta
d\phi \, .
\end{eqnarray} The spherical coordinates are defined in the standard form,
$x=r\sin \theta
\cos
\phi$,
$y=r\sin \theta \sin \phi$ and $z=r \cos \theta$. The angular step is $\Delta
\theta=\Delta \phi=\pi/18$ rad. Plots on the left are contour plots with dark regions
representing the maximum population. Contour lines are in linear scale.
Plots on the right
are surface plots, also in linear scale.

Observe the clear difference between both pairs of plots. For the impact
parameter
$b=1$ a.u., the peak of the ejected electrons is placed along the
positive part of the
$y$-axis ($\theta=\pi/2$ and $\phi=\pi/2$) but also appears a second peak
along the
negative part of the $y$-axis ($\theta=\pi/2$ and $\phi=3\pi/4$). This
second peak is a
consequence of the fact that the ion crosses well over the electronic cloud. The
minimum density of electrons are found along the $x$-axis, the longitudinal direction,
($\theta=\pi/2$, $\phi=0$ and $\theta=\pi/2$, $\phi=\pi$).
However for the impact parameter $b=3$ a.u. only the first
peak can be observed: in this case the ion does not cross over the electronic cloud.

\begin{figure}
\begin{center}
\scalebox{0.75}{\includegraphics{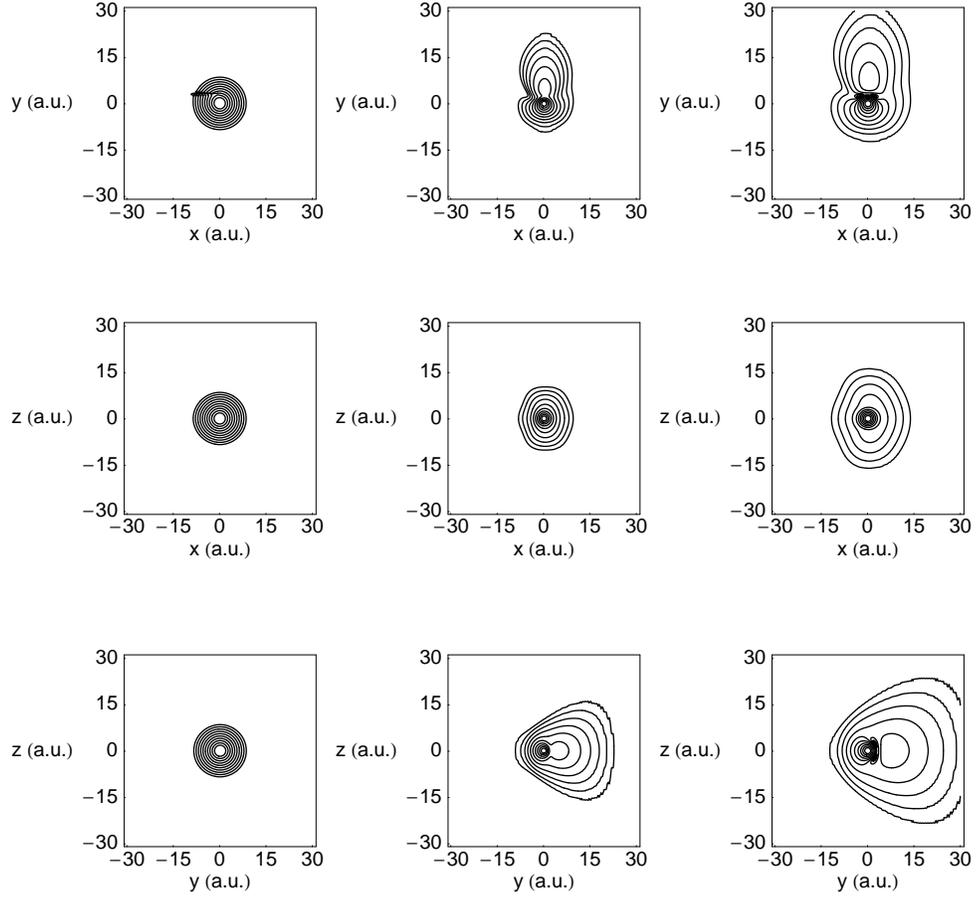}}
\end{center}
\caption{Probability density at different times of the interaction with the
ion, at the
planes $z=0$ (plots at the top), $y=0$ (plots in the middle) and $x=0$
(plots at the
bottom). The impact parameter is $b=3$ a.u. The column on the left
corresponds to the
time at which the ion is placed at the minimum distance to the nucleus
[$x_{ion} (t)=0$]. The
column in the middle is for $x_{ion} (t)=400$ a.u. ($t=11.8$ a.u.) and the
column on the right for
$x_{ion} (t)=600$ a.u. ($t=\tau=15.8$ a.u.). The contour lines are in
logarithmic scale. Atomic units are used.}
\label{fig:cont3}
\end{figure}

Finally, in Fig. \ref{fig:cont3} we plot the probability density at
different times of the
interaction with the ion, at the planes $z=0$ (plots at the top), $y=0$
(plots in the middle)
and $x=0$ (plots at the bottom). The impact parameter is $b=3$ a.u. The
column on the
left corresponds to the time at which the ion is placed at the minimum
distance to the
nucleus [$x_{ion} (t)=0$]. The column in the middle is for $x_{ion} (t)=400$
a.u. ($t=11.8$ a.u.) and the
column on the right for $x_{ion} (t)=600$ a.u. ($t=\tau=15.8$ a.u.). The
contour lines are in
logarithmic scale.
In the frame $z=0$ in the first column, the effect of the incident ion can be
observed as a distortion of the wave-packet. Once the projectile has passed, ionized population moves mainly
along the transverse direction ($y$ axis). The slight trembling in the lower
contour lines that can be appreciated in a few frames, is due to
spatial discretization effects and the non continuous evaluation of the trajectory
of the ion.

\begin{figure}
\begin{center}
\scalebox{0.75}{\includegraphics{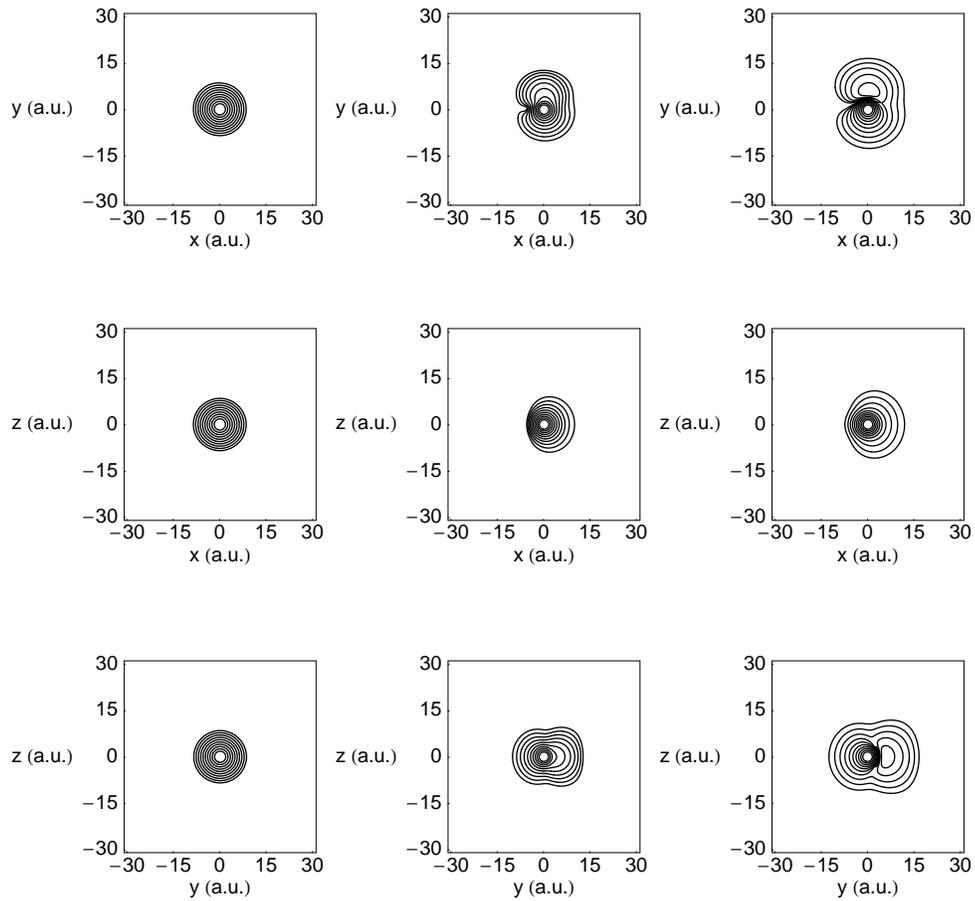}}
\end{center}
\caption{Probability density at different times of the interaction with the
ion for $b=10$ a.u. The arrangement of the plots is the same as in the previous figure.}
\label{fig:cont10}
\end{figure}

The same arrangement of the plots is used in Fig.
\ref{fig:cont10}, but now for the impact parameter $b=10$ a.u.
The distortion of the wave packet is similar to that shown in Fig. \ref{fig:cont3}.
However, ejected electrons are even slower and at the final time the spreading
of the wave-function is not so marked.

\section{Conclusions}

We have presented very realistic three-dimensional numerical simulations to
describe the collision of relativistic heavy ions against hydrogenic atoms.
Our simulations are based on the {\it ab initio} integration of the
time-dependent wave-equation derived from the non-relativistic limit of the
Dirac equation, that is valid for arbitrary electrogmanetic potentials.
This equation differs from the standard time-dependent
Schr\"odinger equation in a term that includes the time derivative
of the scalar potential: neglecting this term leads to a
conceptually wrong formulation.
The recoil of the ion and the motion of the atomic nucleus have been neglected
in our simulations.
We have employed a three-dimensional Cartesian grid to integrate the wave-equation.
As results, we should point out that the probability of excitation of the atom
sharply increases for impact parameters $b<15$ a.u.
However, the ejection of fast electrons is only relevant for even
smaller values of the impact parameter ($b<5$ a.u.).
The space distribution of such electrons is centered along the transversal direction.
It is composed of one maximum (in the positive direction of the $y$ axis) even
for short impact parameters ($b>3$ a.u.). Only in collisions where the ion
crosses over the electronic cloud a more complex structure appears.

\acknowledgements
The authors gratefully acknowledge invaluable discussions with Luis Plaja.
This work has been partially supported by the Spanish Direcci\'on General
de Ense\~nanza
Superior e Investigaci\'on Cient{\'\i}fica (grant PB98-0268) and by the
Junta de Castilla y
Le\'on and Uni\'on Europea, FSE (grant SA044/01).
Computational work was carried out at the Linux cluster of the Optics Group
in the Universidad de Salamanca.

\end{document}